\documentstyle[12pt,thmsa,epsfig,sw20lart]{article}

\input tcilatex
\begin{document}

\title{{\bf Vortices and invariant surfaces generated by symmetries for the 3D
Navier--Stokes equations }}
\author{V. Grassi, R.A. Leo, G. Soliani and P. Tempesta \\
Dipartimento di Fisica dell'Universit\`{a} di Lecce, 73100 Lecce, Italy, \\
and Istituto Nazionale di Fisica Nucleare, Sezione di Lecce, Italy}
\date{}
\maketitle

\begin{abstract}
We show that certain infinitesimal operators of the Lie--point symmetries of
the incompressible 3D Navier--Stokes equations give rise to vortex solutions
with different characteristics. This approach allows an algebraic
classification of vortices and throws light on the alignment mechanism
between the vorticity $\stackrel{\rightarrow}{\omega}$ and the vortex
stretching vector $S\stackrel{\rightarrow}{\omega}$, where $S$ is the strain
matrix. The symmetry algebra associated with the Navier--Stokes equations
turns out to be infinite--dimensional. New vortical structures, generalizing
in some cases well--known configurations such as, for example, the Burgers
and Lundgren solutions, are obtained and discussed in relation to the value
of the dynamic angle $\phi=\arctan\frac{\mid\stackrel{\rightarrow}{\omega}%
\wedge S\stackrel{\rightarrow}{\omega}\mid} {\stackrel{\rightarrow}{\omega}%
\cdot\stackrel{\rightarrow }{\omega}}$. A systematic treatment of the
boundary conditions invariant under the symmetry group of the equations
under study is also performed, and the corresponding invariant surfaces are
recognized.
\end{abstract}

\section{\protect\bigskip Introduction}

Let us consider the incompressible 3D Navier--Stokes equations 
\begin{equation}
\Delta_{1}\equiv\overrightarrow{u_{t}}+\overrightarrow{u}\cdot\nabla 
\overrightarrow{u}+\nabla p-\nu\nabla^{2}\overrightarrow{u}- \overrightarrow{%
f}=0,
\end{equation}
\begin{equation}
\Delta_{2}\equiv\nabla\cdot\overrightarrow{u}=0,
\end{equation}
where $\stackrel{\rightarrow}{u}=(u_{1}(x,y,z,t),u_{2}(x,y,z,t),
u_{3}(x,y,z,t))$ is the velocity field, $p=p(x,y,z,t)$ the fluid pressure, $%
\nu $ the viscosity coefficient, and $\stackrel{\rightarrow}{f}=\stackrel{%
\rightarrow}{f}(x,y,z,t)$ is an external force.

Generally, in the studies of turbulence flows it is important to recognize
the generation mechanisms and evolution of vortical structures. In this task
both analytical and numerical methods are of help. For example, Tanaka and
Kida [1] showed that high vorticity with relatively low strain rate
corresponds to the vortex tube, while high vorticity with comparable strain
rate to the vortex sheet. Wray and Hunt [2] presented an algorhitm for
classification of turbulent structures where the flow field is classified
into four regions (eddies, convergence, shear and stream regions).

Interesting results for the 3D Navier--Stokes equations, adressed to this
line of research, have been found recently by Galanti, Gibbon and Heritage
(GGH) [3] and by Gibbon, Fokas and Doering (GFD) [4]. The first authors
investigated the mechanism of the vorticity alignment in the Navier--Stokes
isotropic turbulence. They started from the results obtained in [5],
according to which the vorticity vector $\stackrel{\rightarrow }{\omega }$
aligns with the intermediate eigenvector of the strain matrix $S$. This
problem was tackled via the introduction of the variables $\alpha =\frac{%
\stackrel{\rightarrow }{\omega }\cdot S\stackrel{\rightarrow }{\omega }}{%
\stackrel{\rightarrow }{\omega }\cdot \stackrel{\rightarrow }{\omega }}%
\equiv $ $\stackrel{\wedge }{\xi }\cdot $ $S\stackrel{\wedge }{\xi }$ (the
''stretching rate'') and $\stackrel{\rightarrow }{\chi }=\frac{\stackrel{%
\rightarrow }{\omega }\wedge S\stackrel{\rightarrow }{\omega }}{\stackrel{%
\rightarrow }{\omega }\cdot \stackrel{\rightarrow }{\omega }}\stackrel{%
\wedge }{\equiv \xi }\wedge S\stackrel{\wedge }{\xi }$ where $\stackrel{%
\wedge }{\xi }$ $=\frac{\stackrel{\rightarrow }{\omega }}{\omega }$. Two
differential equations were derived for $\alpha $ and $\chi $ which were
exploited to discuss the vorticity alignment in terms of the angle $\phi
(x,y,z,t)$ between $\stackrel{\rightarrow }{\omega }$ and $S\stackrel{%
\rightarrow }{\omega }$ defined by tan $\phi =\frac{\mid \stackrel{%
\rightarrow }{\omega }\wedge S\stackrel{\rightarrow }{\omega }\mid }{%
\stackrel{\rightarrow }{\omega }\cdot \stackrel{\rightarrow }{S\omega }}=%
\frac{\chi }{\alpha }$ $(\chi =\sqrt{\stackrel{\rightarrow }{\chi }\cdot 
\stackrel{\rightarrow }{\chi }})$. On the other hand, GFD observed that
stretched vortex solutions of the 3D Navier--Stokes equations, such as the
Burgers vortices, are characterized by a uni--directional vorticity
stretched by a strain field decoupled from them. Then, they showed that
these drawbacks can be partially circumvented by searching classes of
solutions of the type $\stackrel{\rightarrow }{u}%
=(u_{1}(x,y,t),u_{2}(x,y,t),\gamma (x,y,t)z+W(x,y,t))$. In such a way the
equations for the third component of vorticity $\omega _{3}$ and $W$
decouple. Generalizations of the Burgers vortex type are constructed and
various solutions for $W$ are discussed.

Following the spirit of the above mentioned works, here we show as vortex
solutions of the incompressible 3D Navier--Stokes equations (1.1)--(1.2) are
generated by their symmetry infinitesimal operators. Using a
group--theoretical approach, new classes of exact solutions are found. Some
of them contain as special cases the Burgers vortex and the shear-layer
solutions, which are produced by two different infinitesimal operators. By
an algebraic point of view, equations (1.1)--(1.2) are characterized by a
symmetry algebra containing arbitrary functions. This fact indicates that
this algebra is infinite--dimensional. Subalgebras isomorphic to the algebra
of the Euclidean group are present.

An important feature of the technique applied is that the symmetry algebra
turns out to be not trivial only if the external force $\stackrel{%
\rightarrow }{f}$ is derivable from a potential function $\varphi (x,y,z,t)$%
, i.e. 
\begin{equation}
\stackrel{\rightarrow }{f}=\nabla \varphi .
\end{equation}
The problem of the boundary conditions is also considered. This is a
fundamental aspect of the symmetry analysis, because if the boundary
conditions are also invariant under the symmetry group, then an invariant
solution is the unique solution of the system under investigation [6]. In
this way, we have found interesting invariant surfaces which play the role
of invariant conditions.

The paper is organized as follows. In Section 2 the infinitesimal operators
of the (Lie--point) symmetries admitted by Eqs. (1.1)--(1.2) are obtained,
while the corresponding group transformations are described in Section 3.
Section 4 is devoted to the determination of the reduced equations coming
from the generators of the symmetry subgroups. Some of these equations can
be exactly solved and give rise to solutions of the vortex type which
generalize well--known configurations as, for example, the Burgers
shear--layer and the Burgers vortex. Section 5 deals with the invariance
properties of the boundary conditions. Finally, in Section 6 and in
Appendices A and B some comments and details of the calculations are
reported, respectively.

\section{Group analysis}

\setcounter{equation}{0}

To the aim of finding the symmetry group of Eqs. (1.1)--(1.2), let us
introduce the vector field

\begin{equation}
V=\xi_{1}\partial_{x}+\xi_{2}\partial_{y}+\xi_{3}\partial_{z}+\xi_{4}
\partial_{t}+\stackrel{}{\stackrel{3}{\stackunder{j=1}{\sum}}}\phi_{j}
\partial_{u_{j}}+\phi_{4}\partial_{p},
\end{equation}
where $\xi_{j}$ and $\phi_{j}$ are functions of $x,y,z,t,u_1,u_2,u_3,p$, and 
$\partial_{x}=\partial /\partial x$, and so on.

We remind the reader that for a given system of differential equations $%
\Delta _{j}=0$, defined on a differentiable manifold $M$, a local group of
transformations $G$, acting on $M$, is a group of symmetries of $\Delta
_{j}=0$ if and only if 
\begin{equation}
pr^{(n)}V[\Delta _{j}]=0
\end{equation}
whenever $\Delta _{j}=0,$ for every generator $V$ of $G$ (see [7], [8] and
[9]).

In the case of Eqs. (1.1)--(1.2), the condition (2.2) becomes \smallskip 
\begin{equation}
pr^{(2)}V[\Delta_{j}]=0,
\end{equation}
where $pr^{(2)}V$ denotes the second prolongation

\[
pr^{(2)}V=V+\phi _{1}^{x}\partial _{u_{1x}}+\phi _{1}^{y}\partial
_{u_{1}y}+\phi _{1}^{z}\partial _{u_{1z}}+\phi _{1}^{t}\partial
_{u_{1t}}+\phi _{1}^{xx}\partial _{u_{1xx}}+\phi _{1}^{yy}\partial
_{u_{1yy}}+ 
\]
\[
\phi _{1}^{zz}\partial _{u_{1zz}}+\phi _{2}^{x}\partial _{u_{2x}}+\phi
_{2}^{y}\partial _{u_{2y}}+\phi _{2}^{z}\partial _{u_{2z}}+\phi
_{2}^{t}\partial _{u_{2t}}+\phi _{2}^{xx}\partial _{u_{2xx}}+\phi
_{2}^{yy}\partial _{u_{2yy}}+ 
\]
\[
\phi _{2}^{zz}\partial _{u_{2zz}}+\phi _{3}^{x}\partial _{u_{3x}}+\phi
_{3}^{y}\partial _{u_{3y}}+\phi _{3}^{z}\partial _{u_{3z}}+\phi
_{3}^{t}\partial _{u_{3t}}+\phi _{3}^{xx}\partial _{u_{3xx}}+\phi
_{3}^{yy}\partial _{u_{3yy}}+ 
\]
\begin{equation}
\phi _{3}^{zz}\partial _{u_{3zz}}+\phi _{4}^{x}\partial _{p_{x}}+\phi
_{4}^{y}\partial _{p_{y}}+\phi _{4}^{z}\partial _{p_{z}},
\end{equation}
with 
\begin{equation}
\phi _{j}^{\alpha }=D_{\alpha }(\phi _{j}-\stackrel{4}{\stackunder{i=1}{\sum 
}}\xi _{i}u_{i}^{j})+\stackrel{4}{\stackunder{i=1}{\sum }}\xi _{i}u_{\alpha
,i}^{j},
\end{equation}
$j=1,2,3,4$, $u_{i}^{j}=\frac{\partial u_{j}}{\partial x_{i}}$, $u_{\alpha
,i}^{j}=\frac{\partial u_{\alpha}^{j}}{\partial x_{i}}$, $x_{i}$ and $\alpha$
stand for $x$, $y$, $z$, $t$, and $x$, $y$, $z$, $t$, $xx$, $yy$, $zz$,
respectively, $D$ being the derivation operation.

Equation (2.3) provides a set of differential equations, the so--called {\it 
{determining system}}, which allows us to obtain the coefficients $%
\xi_{j},\phi_{j}$. From these we infer the infinitesimal operator generating
the symmetry group of Eqs. (1.1)--(1.2), namely

$\;$%
\begin{equation}
V=\stackrel{9}{\stackunder{i=1}{\sum }}V_{i},
\end{equation}
where 
\begin{equation}
V_{1}=g\partial _{x}+\stackrel{.}{g}\partial _{u_{1}}+(-\stackrel{..}{g}%
x+gf_{1})\partial _{p},
\end{equation}

\begin{equation}
V_{2}=h\partial _{y}+\stackrel{.}{h}\partial _{u_{2}}+(-\stackrel{..}{h}%
y+hf_{2})\partial _{p},
\end{equation}

\begin{equation}
V_{3}=r\partial _{z}+\stackrel{.}{r}\partial _{u_{3}}+(-\stackrel{..}{r}%
z+rf_{3})\partial _{p},
\end{equation}

\begin{equation}
V_{4}=k\partial _{p},
\end{equation}

\[
V_{5}=a[x\partial _{x}+y\partial _{y}+z\partial _{z}+2t\partial _{t}-%
\stackrel{3}{\stackunder{i=1}{\sum }}u_{i}\partial _{u_{i}} 
\]

\begin{equation}
+(-2p+xf_{1}+yf_{2}+zf_{3}+2\int^{x}f_{1}dx+2t\int^{x}f_{1t}dx)\partial
_{p}],
\end{equation}

\begin{equation}
V_{6}=b[y\partial _{x}-x\partial _{y}+u_{2}\partial _{u_{1}}-u_{1}\partial
_{u_{2}}+(yf_{1}-xf_{2})\partial _{p}],
\end{equation}

\begin{equation}
V_{7}=c[z\partial _{y}-y\partial _{z}+u_{3}\partial _{u_{2}}-u_{2}\partial
_{u_{3}}+(zf_{2}-yf_{3})\partial _{p}],
\end{equation}

\begin{equation}
V_{8}=d[z\partial _{x}-x\partial _{z}+u_{3}\partial _{u_{1}}-u_{1}\partial
_{u_{3}}+(zf_{1}-xf_{3})\partial _{p}],
\end{equation}

\begin{equation}
V_{9}=e[\partial _{t}+(\int^{x}f_{1t}dx)\partial _{p}],
\end{equation}
\noindent $g(t)$, $h(t)$, $k(t)$, $R(t)$ are arbitrary functions of time of
class $C^{\infty}$, and $a$, $b$, $c$, $d$, $e$ arbitrary constants (dots
mean time derivatives). The infinitesimal operators (2.7)--(2.15) represent
the generators of the Lie--point symmetries of the 3D Navier--Stokes
equations (1.1)--(1.2).

\smallskip

\section{Group transformations}

\setcounter{equation}{0}

The integration of the infinitesimal operator (2.6) enables us to find the
finite transformations leaving the equations (1.1)--(1.2) invariant. We have
that the linear combination $W=V_{1}+V_{2}+V_{3}$ gives rise to a
transformation to an arbitrary moving coordinate frame of the type 
\begin{equation}
G_{\stackrel{\rightarrow}{\alpha}}:(\stackrel{\rightarrow}{x},t,\stackrel{%
\rightarrow}{u},p)\longrightarrow(\stackrel{\rightarrow}{x}+\varepsilon 
\stackrel{\rightarrow}{\alpha},t,\stackrel{\rightarrow}{u}+\varepsilon 
\stackrel{\rightarrow}{\alpha},p-\varepsilon\stackrel{\rightarrow}{x} \cdot%
\stackrel{..}{\stackrel{\rightarrow}{\alpha}}-\frac{1}{2}\varepsilon ^{2}%
\stackrel{\rightarrow}{\alpha}\cdot\stackrel{..}{\stackrel{\rightarrow}{%
\alpha}}+\int_{0}^{\varepsilon}\stackrel{.}{\stackrel{\rightarrow}{\alpha}}%
\cdot\stackrel{\rightarrow}{f}d\varepsilon)
\end{equation}
where $\stackrel{}{\overrightarrow{\alpha}}=(g(t),h(t),r(t))\ {\rm {and}\
\varepsilon}$ is the group parameter.

$V_{4}$ is the generator of the pressure change

\smallskip

\begin{equation}
G_{p}\text{ }:\text{ }(\stackrel{\rightarrow}{x},t,\stackrel{\rightarrow}{u}%
,p)\longrightarrow \text{ }(\stackrel{\rightarrow}{x},t, \stackrel{%
\rightarrow}{u},p+\varepsilon\stackrel{}{k(t)})\text{ ,}
\end{equation}

\smallskip

\noindent while $V_{5}$ is the infinitesimal operator for the scale
transformations. The vector fields $V_{6},V_{7},V_{8}$ yield both the space
and the velocity field rotations. Finally, $V_{9}$ produces the time
translations together with an adjustment of the pressure, i.e.

\smallskip 
\begin{equation}
G:(\stackrel{\rightarrow }{x},t,\stackrel{\rightarrow }{u},p)\longrightarrow
\left( \stackrel{\rightarrow }{x},\,t+\varepsilon ,\,\stackrel{\rightarrow }{%
u},\,\,p+\varphi (x,y,z,t+\varepsilon )-\varphi (x,y,z,t)\right) ,
\end{equation}

\bigskip\noindent where $\varphi \,\,$is the potential function linked to
the external force by (1.3). Then, we can ''absorbe'' $\stackrel{\rightarrow%
}{f} $ in $p$ by re--defining the pression $p$ as 
\[
p^{\prime }=p-\varphi (x,y,z,t). 
\]
Without introducing new symbols, we shall put formally $\stackrel{\rightarrow%
}{f}=0$ into Eqs. (1.1)--(1.2) and into the expressions (2.7)--(2.15) for
the generators.

\section{Reduced equations and exact solutions}

\setcounter{equation}{0}

By exploiting the generators $V_{j}$ of the Lie--point transformations (see
(2.7)--(2.15)), one can build up exact solutions of Eqs. (1.1)--(1.2) via
the symmetry reduction approach. This allows one to lower the order of the
system of differential equations under consideration using the invariants
associated with a given subgroup of the symmetry group. In the following we
present some reductions leading to exact solutions of Eqs. (1.1)--(1.2) of
physical interest.

\smallskip

\subsection{Case i)}

Let us take the vector field $V_{1}$ defined by (2.7). A set of basis
invariants $I_{j}$ of the related subgroup can be determined from the finite
group transformations

\begin{eqnarray}
x^{^{\prime }} &=&x+\varepsilon g(t),\text{ }y^{\prime }=y,z^{\prime
}=z,t^{\prime }=t,  \nonumber \\
u_{1}^{\prime } &=&u_{1}+\varepsilon \stackrel{.}{g},u_{2}^{\prime
}=u_{2},u_{3}^{\prime }=u_{3},  \nonumber \\
p^{\prime } &=&p-\frac{1}{2}\varepsilon \stackrel{..}{g}(2x+\varepsilon g).
\end{eqnarray}

\smallskip \noindent We obtain 
\begin{equation}
I_{1}=y,I_{2}=z,I_{3}=t,I_{4}=u_{1}-\frac{\stackrel{.}{g}}{g}%
x,I_{5}=u_{2},I_{6}=u_{3},I_{7}=p+\frac{\stackrel{..}{g}}{2g}x^{2}.
\end{equation}
By means of the choice of variables 
\begin{eqnarray}
U_{1}(y,z,t) &\equiv &I_{4}=u_{1}-\frac{\stackrel{.}{g}}{g}
x,U_{2}(y,z,t)\equiv I_{5}=u_{2},U_{3}(y,z,t)\equiv u_{3},  \nonumber \\
\pi (y,z,t) &\equiv &I_{7}=p+\frac{\stackrel{..}{g}}{2g}x^{2},
\end{eqnarray}
the system (1.1)--(1.2) is cast into the reduced form

\smallskip

\begin{equation}
U_{1t}+\frac{\stackrel{.}{g}}{g}U_{1}+U_{2}U_{1y}+U_{3}U_{1z}-\nu
(U_{1yy}+U_{1zz})=0,  \tag{4.4a}
\end{equation}

\begin{equation}
U_{2t}+U_{2}U_{2y}+U_{3}U_{2z}+\pi _{y}-\nu (U_{2yy}+U_{2zz})=0,  \tag{4.4b}
\end{equation}

\begin{equation}
U_{3t}+U_{2}U_{3y}+U_{3}U_{3z}+\pi _{z}-\nu (U_{3yy}+U_{3zz})=0,  \tag{4.4c}
\end{equation}

\begin{equation}
U_{2y}+U_{3z}+\frac{\stackrel{.}{g}}{g}=0.  \tag{4.4d}
\end{equation}

A particular solution of Eqs. (4.4a)--(4.4d) can be found as follows. Let us
put [10] 
\begin{equation}
U_{2}=k_{1}y,U_{3}=\sigma -k_{1}z,\pi =-k_{1}^{2}\frac{z^{2}}{2}+k_{1}\sigma
z-k_{1}^{2}\frac{y^{2}}{2}+\tau (t),  \tag{4.5}
\end{equation}
where $k_{1},k_{2},\sigma ,g$ are constants, and $\tau (t)$ is an arbitrary
function of time. Substitution from (4.5) into Eqs. (4.4a)--(4.4d) yields
the linear equation

\smallskip 
\begin{equation}
U_{1t}+k_{1}yU_{1y}+(\sigma -k_{1}z)U_{1z}-\nu (U_{1yy}+U_{1zz})=0. 
\tag{4.6}
\end{equation}

At this point it is convenient to look for a solution of Eq. (4.6) of the
form 
\begin{equation}
U_{1}=Y(y)\,T(z)\,\Phi \left( t\right) .  \tag{4.7}
\end{equation}
Then, Eq. (4.6) can be written as 
\begin{equation}
k_{1}y\frac{Y_{y}}{Y}-\nu \frac{Y_{yy}}{Y}=(k_{1}z-\sigma )\frac{T_{z}}{T}
+\nu \frac{T_{zz}}{T}-\frac{\Phi _{t}}{\Phi }\equiv G,  \tag{4.8}
\end{equation}
$G$ being an arbitrary constant. On the other hand, Eq. (4.8) tells us that

\smallskip

\[
(k_{1}z-\sigma )\frac{T_{z}}{T}+\nu \frac{T_{zz}}{T}=G+\frac{\Phi _{t}}{\Phi 
}\equiv H, 
\]

\noindent where $H$ is an arbitrary constant. To summarize, the functions $Y$%
, $T$, $\Phi$ obey the ordinary differential equations 
\begin{equation}
\Phi _{t}=(H-G)\,\Phi ,  \tag{4.9a}
\end{equation}

\begin{equation}
\nu Y_{yy}-k_{1}yY_{y}+GY=0,  \tag{4.9b}
\end{equation}

\begin{equation}
\nu T_{zz}+(k_{1}z-\sigma )T_{z}-HT=0.  \tag{4.9c}
\end{equation}

\noindent Equations (4.9a)--(4.9c) afford, respectively, the general
solutions

\begin{equation}
\Phi =c_{1}\exp [(H-G)t],  \tag{4.10a}
\end{equation}

\begin{equation}
Y=c_{2}M(-\frac{G}{2k_{1}},\frac{1}{2},\frac{k_{1}y^{2}}{2\nu })+yc_{3}M( 
\frac{1}{2}-\frac{G}{2k_{1}},\frac{3}{2},\frac{k_{1}y^{2}}{2\nu }), 
\tag{4.10b}
\end{equation}

\begin{equation}
T=c_{4}M(-\frac{H}{2k_{1}},\frac{1}{2},\frac{-k_{1}(z-\frac{\sigma }{k_{1}}
)^{2}}{2\nu })+(z-\frac{\sigma }{k_{1}})c_{5}M(\frac{1}{2}-\frac{H}{2k_{1}}, 
\frac{3}{2},\frac{-k_{1}(z-\frac{\sigma }{k_{1}})^{2}}{2\nu }),  \tag{4.10c}
\end{equation}
with $c_{1},...,c_{5}$ arbitrary constants. $M$ is the Kummer function
defined by \cite{7} 
\begin{equation}
M(\alpha ,\beta ,z)=1+\frac{\alpha }{\beta }z+\frac{(\alpha )_{2}}{(\beta
)_{2}}\frac{z^{2}}{2!}+...+\frac{(\alpha )_{n}}{(\beta )_{n}}\frac{z^{n}}{n!}%
+...,  \nonumber
\end{equation}
with 
\begin{equation}
(\alpha )_{n}=\alpha (\alpha +1)(\alpha +2)...(\alpha +n-1),\text{ }(\alpha
)_{0}=1.  \nonumber
\end{equation}
Consequently, from (4.3) we obtain 
\begin{eqnarray}
&&u_{1} =c_{1}e^{(H-G)t}\{[c_{2}M(-\frac{G}{2k_{1}},\frac{1}{2},\frac{
k_{1}y^{2}}{2\nu })+yc_{3}M(\frac{1}{2}-\frac{G}{2k_{1}},\frac{3}{2},\frac{
k_{1}y^{2}}{2\nu })]\times  \nonumber \\
&&[c_{4}M(-\frac{H}{2k_{1}},\frac{1}{2},\frac{-k_{1}(z-\frac{\sigma }{k_{1}}
)^{2}}{2\nu })+(z-\frac{\sigma }{k_{1}})c_{5}M(\frac{1}{2}-\frac{H}{2k_{1}}, 
\frac{3}{2},\frac{-k_{1}(z-\frac{\sigma }{k_{1}})^{2}}{2\nu })]\},  \nonumber
\end{eqnarray}
{\hspace{12.5cm}(4.11a)} \smallskip 
\begin{equation}
u_{2}=k_{1}y,  \tag{4.11b}
\end{equation}

\begin{equation}
u_{3}=\sigma -k_{1}z,  \tag{4.11c}
\end{equation}

\begin{equation}
p=-k_{1}^{2}\frac{z^{2}}{2}+k_{1}\sigma z+\tau (t)-k_{1}^{2}\frac{y^{2}}{2}.
\tag{4.11d}
\end{equation}
It is noteworthy that at least in some special cases, Eqs. (4.11a)--(4.11d)
reproduce interesting solutions which can be interpreted by a physical point
of view, such as the Burgers shear-layer and other solutions. This aspect is
discussed below.

\subsubsection{\protect\bigskip The Burgers shear-layer and other solutions}

\smallskip We observe that Eqs. (4.11a)--(4.11d) give rise to static
solutions for $H=G\equiv \lambda $. We shall distinguish two cases: $a)$ $%
\lambda =0$, and $b)$ $\lambda \neq 0$.

$\ $

$Case$ $a)$

From (4.1a) we obtain

\begin{equation}
u_{1}=[c_{2}+c_{3}\sqrt{-\frac{\pi \nu }{k_{1}}}\func{erf}(\sqrt{-\frac{k_{1}%
}{2\nu }}y)][c_{4}+c_{5}\sqrt{\frac{\pi \nu }{k_{1}}}\func{erf}(\sqrt{\frac{%
k_{1}}{2\nu }}z)],  \tag{4.12}
\end{equation}
where for simplicity $\sigma \equiv 0$ and we have used the properties $%
M(0,b,\varsigma )=1$ and \bigskip [11, p. 509]]

\[
M(\frac{1}{2},\frac{3}{2},-\varsigma ^{2})=\frac{\sqrt{\pi }}{2\varsigma} 
\func{erf}\text{ }\varsigma , 
\]
with 
\[
\func{erf}\text{ }\varsigma =\frac{2}{\sqrt{\pi }}\int_{0}^{\varsigma}
\exp(-\theta^{2})d\theta 
\]

\bigskip We remark that by choosing $k_{1}=-\gamma $ $(\gamma >0)$ and $%
c_{5}=0$ in Eq.(4.12), we get the Burgers shear--layer solution \cite{3,1}

\begin{equation}
\stackrel{\rightarrow }{u}=(u_{1}(y),-\gamma y,\gamma z)^{T},  \tag{4.13}
\end{equation}
where 
\begin{equation}
Y(y)\equiv u_{1}(y)=A\sqrt{\frac{\pi\nu}{\gamma}}\func{erf} (\sqrt{\frac{%
\gamma}{2\nu}}y)+B,  \tag{4.14}
\end{equation}
with A and B arbitrary constants.

\bigskip Now let us recall some basic quantities appearing in the study of
turbulent flows. They are the vorticity $\stackrel{\rightarrow }{\omega }%
=\nabla \wedge \stackrel{\rightarrow }{u}$, the strain matrix $S$ whose
elements are defined by $S_{ij}=\frac{1}{2}(u_{i,j}+u_{j,i}),$ where $%
u_{i,j}\equiv \frac{\partial u_{i}}{\partial x_{j}}$ and\ $u_{j,i}\equiv 
\frac{\partial u_{j}}{\partial x_{i}},$ the energy dissipation $S_{ij}S_{ij}$
and the enstrophy $\omega _{i}\omega _{i}$ (the summation under repeated
indices is understood) [11]--[12]. Furthermore, in the study of the vortex
alignments, it is important the dynamic angle $\phi (x,y,z,t)$ defined in
the Introduction.

The vorticity corresponding to the solution $(4.14)$ is $\stackrel{%
\rightarrow }{\omega }=(0,0,\omega _{3})^{T}$ with 
\begin{equation}
\omega _{3}=-Y^{\prime }(y)=-A\sqrt{\frac{2\nu }{\gamma }}\exp (-\frac{%
\gamma y^{2}}{2\nu }),  \tag{4.15}
\end{equation}

\bigskip \noindent while the strain matrix $S$ is given by

\begin{equation}
S=\left( 
\begin{array}{ccc}
0 & \frac{1}{2}Y^{\prime } & 0 \\ 
\frac{1}{2}Y^{\prime } & -\gamma & 0 \\ 
0 & 0 & \gamma
\end{array}
\right) {\Huge .}  \tag{4.16}
\end{equation}
The energy dissipation reads 
\begin{equation}
S_{ij}S_{ij}=2\gamma ^{2}+\frac{1}{2}Y^{\prime 2},  \tag{4.17}
\end{equation}
which tells us that the dissipation is high when the vorticiy is so [4]. We
observe that the vorticity vanishes for large values of $y^{2}$. We notice
also that the pression $p$ satisfies the equation 
\[
\frac{\partial p}{\partial z}=-\gamma z, 
\]
which is just the equation (66) shown in [3]. Within our framework, $p$
arises as a consequence of the reduction procedure involving the generator $%
V_{1}$. The dynamic angle $\phi $ related to the vortex solution (4.13)
turns out to be zero, so that the vectors $\stackrel{\rightarrow }{\omega }$
and $S\stackrel{\rightarrow }{\omega }$ are parallel.

From formula (4.12) we deduce that for $k_{1}>0,$ by switching off the
constant $c_{3}$ a shear-layer of the Burgers type of the form $\stackrel{%
\rightarrow }{u}=(Z(z),-\gamma y,\gamma z)^{T}$ emerges, where now

\smallskip 
\begin{equation}
Z(z)\equiv u_{1}(z)=A\sqrt{\frac{\pi \nu }{k_{1}}}\func{erf}(\sqrt{\frac{%
k_{1}}{2\nu }}z)+B.  \tag{4.18}
\end{equation}
Therefore, (4.12) can be interpreted as a more general vortex configuration
of the Burgers shear-layer type in which both the variables $y$ and $z$ are
involved. \smallskip $Case$ $b)$

Formula (4.11a) can be exploited to find other interesting static solutions
which can be considered as generalizations of the Burgers shear--layer. An
explicit example corresponds to the choice $\lambda =-k_{1}$ and $%
\sigma=c_{3}=c_{5}=0$. We have 
\begin{equation}
u_{1}=A\exp [\frac{k_{1}}{2\nu }(y^{2}-z^{2})],  \tag{4.19a}
\end{equation}

\begin{equation}
u_{2}=k_{1}y,  \tag{4.19b}
\end{equation}

\begin{equation}
u_{3}=-k_{1}z,  \tag{4.19c}
\end{equation}
\noindent $A$ being a constant.

The vorticity $\stackrel{\rightarrow}{\omega}$ has $\omega_{1}=0$ and the
remaining components different from zero. Precisely

\begin{equation}
\omega _{2}=-\frac{Ak_{1}}{\nu }z\exp [\frac{k_{1}}{2\nu }(y^{2}-z^{2})], 
\tag{4.20a}
\end{equation}

\begin{equation}
\omega _{3}=-\frac{Ak_{1}}{\nu }y\exp [\frac{k_{1}}{2\nu }(y^{2}-z^{2})]. 
\tag{4.20b}
\end{equation}

\bigskip

A possible physical interpretation of this situation is that for $k_{1}<0,$
it represents a vortex structure where $\omega_{2}$ behaves as a
shear--layer of the Burgers type along the axis $y$ for any fixed value of $%
z $, while in opposition to $\omega _{2}$, $\omega _{3}$ vanishes at $y=0$.
The evaluation of the dynamic angle for the vortex (4.19a)--(4.19c) leads to
the formula

\begin{equation}
\tan \phi =\frac{2xy}{y^{2}-z^{2}}\equiv \sinh 2\theta ,  \tag{4.21}
\end{equation}
\noindent where we have put $y=\cosh \theta $, $z=\sinh \theta $. For
example, we have $\phi =0$ for $\theta =0$, and $\phi =\pm \frac{\pi }{2}$
for $\theta \rightarrow \pm \infty $.

\bigskip

{\it An example of a non--static solution}

\bigskip

Non--static solutions can be obtained from (4.11a)--(4.11d) assuming that $%
H-G\neq 0$. An interesting example is given by

\begin{equation}
u_{1}=A\exp (-\frac{\gamma }{2}t)\exp (-\frac{\gamma y^{2}}{4\nu })y^ {\frac{%
1}{2}}I_{-\frac{1}{4}}(\frac{\gamma y^{2}}{4\nu}),  \tag{4.22a}
\end{equation}

\begin{equation}
u_{2}=-\gamma y,  \tag{4.22b}
\end{equation}

\begin{equation}
u_{3}=\gamma z,  \tag{4.22c}
\end{equation}

\begin{equation}
p=-\frac{\gamma ^{2}}{2}(y^{2}+z^{2}),  \tag{4.22d}
\end{equation}
which is derived from (4.11a)--(4.11d) for $H=0$,$G=\frac{\gamma }{2}
,k_{1}=-\gamma (\gamma >0),c_{3}=0$, $c_{5}=0,\sigma=0$, $\tau =0$, where $I$
denotes the modified Bessel function.

The vorticity is $\stackrel{\rightarrow }{\omega }\,=(0,0,\omega _{3}),$
with 
\begin{equation}
\omega _{3}=-u_{1y}=B\exp (-\frac{\gamma }{2}t)\exp (-\frac{\gamma y^{2}}{%
4\nu })y^{\frac{3}{2}}[I_{-\frac{1}{4}}(\frac{\gamma y^{2}}{4\nu })-I_{\frac{%
3}{4}}(\frac{\gamma y^{2}}{4\nu })].  \tag{4.23}
\end{equation}

The strain matrix and the dissipation corresponding to the solution
(4.22a)--(4.22d) can be derived from (4.16) and (4.17), respectively, via
the substitution of $Y^{\prime }$ by $u_{1y}$ (see (4.13)). When $\nu$ and $%
\gamma$ are fixed, for large $y$ we have

\[
\omega _{3}\sim y^{-\frac{3}{2}}, 
\]

\noindent while the total strain behaves as

\begin{equation}
S_{ij}S_{ij}\sim 2\gamma^{2}+\frac{A^{2}\nu}{8\pi\gamma}y^{-3}.  \tag{4.24}
\end{equation}
The dynamic angle corresponding to the solution (4.40)-(4.42) is zero.
Therefore, this vortex-like structure is characterized by a vorticity which
is aligned with $S\stackrel{\rightarrow }{\omega }.$

\subsection{\bf Case ii)}

The vector field

\begin{equation}
V_{2}=h\partial _{y}+\stackrel{.}{h}\partial _{u_{2}}-\stackrel{}{\stackrel{%
..}{h}}y\,\partial _{p}  \tag{4.25}
\end{equation}
is related to the invariants

\begin{equation}
t,x,z,u_{2}-\frac{\stackrel{.}{h}}{h}\,y,u_{1},u_{3}.  \tag{4.26}
\end{equation}
Then, by using the variables

\begin{equation}
u_{1}=U_{1}(x,z,t),  \tag{4.27a}
\end{equation}
\begin{equation}
u_{2}=U_{2}(x,z,t)+\frac{\stackrel{.}{h}}{h}y,  \tag{4.27b}
\end{equation}

\begin{equation}
u_{3}=U_{3}(x,z,t),  \tag{4.27c}
\end{equation}

\begin{equation}
p=\pi -\frac{\stackrel{..}{h}}{2h}y^{2},  \tag{4.27d}
\end{equation}

\noindent we are led to the reduced system

\begin{equation}
U_{1t}+U_{1}U_{1x}+U_{3}U_{1z}+\pi _{x}-\nu (U_{1xx}+U_{1zz})=0,  \tag{4.28a}
\end{equation}
\begin{equation}
U_{2t}+U_{1}U_{2x}+U_{2}\frac{\stackrel{.}{h}}{h}+U_{3}U_{2z}-\nu
(U_{2xx}+U_{2zz})=0,  \tag{4.28b}
\end{equation}
\begin{equation}
U_{3t}+U_{1}U_{3x}+U_{3}U_{3z}+\pi _{z}-\nu (U_{3xx}+U_{3zz})=0,  \tag{4.28c}
\end{equation}
\begin{equation}
U_{1x}+U_{3z}+\frac{\stackrel{.}{h}}{h}=0.  \tag{4.28d}
\end{equation}
By changing $x$ with $y$ and $U_{2}$ with $U_{1},$ this system can be
discussed in a way similar to that followed for Eqs. (4.4a)--(4.4d). Of
course, starting from the field $V_{3}$ a reduced system analogous to Eqs.
(4.28a)--(4.28d) is derived.

\smallskip

\subsection{\bf Case iii)}

Let us deal with the vector field

\[
W=V_{1}+V_{2}+V_{3}=g\partial _{x}+h\partial _{y}+r\partial _{z}+\stackrel{.%
}{g}\partial _{u_{1}}+\stackrel{.}{h}\partial _{u_{2}}+\stackrel{.}{r}%
\partial _{u_{3}} 
\]
\begin{equation}
+(\stackrel{..}{g}x+\stackrel{..}{h}y+\stackrel{..}{r}z)\partial _{p}. 
\tag{4.29}
\end{equation}
The group transformations associated with $W$ are

\begin{eqnarray}
x^{\prime } &=&x+\varepsilon g,y^{\prime }=y+\varepsilon h,z^{\prime
}=z+\varepsilon r,t^{\prime }=t,  \nonumber \\
u_{1}^{\prime } &=&u_{1}+\varepsilon \stackrel{.}{g},u_{2}^{\prime
}=u_{2}+\varepsilon \stackrel{.}{h},u_{3}^{\prime }=u_{3}+\varepsilon 
\stackrel{.}{r},  \nonumber \\
p^{\prime } &=&p-\frac{\varepsilon \stackrel{..}{g}}{2}(2x+\varepsilon g)-%
\frac{\varepsilon \stackrel{..}{h}}{2}(2y+\varepsilon h)-\frac{\varepsilon 
\stackrel{..}{r}}{2}(2z+\varepsilon r)  \tag{4.30}
\end{eqnarray}
From (4.30) we deduce the invariants

\begin{equation}
I_{o}=t,I_{1}=u_{1}-x\frac{\stackrel{.}{g}}{g},I_{2}=u_{2}-y\frac{\stackrel{.%
}{h}}{h},I_{3}=z\,\frac{\stackrel{.}{r}}{r},  \nonumber
\end{equation}
\begin{equation}
I_{4}=p+\frac{1}{2}x^{2}\frac{\stackrel{.}{g}}{g}+\frac{1}{2}y^{2}\frac{%
\stackrel{.}{h}}{h}+\frac{1}{2}z^{2}\frac{\stackrel{.}{r}}{r}.  \tag{4.31}
\end{equation}

Now let us introduce the variables

\begin{equation}
U_{1}=I_{1},U_{2}=I_{2},U_{3}=I_{3},\pi =I_{4},  \tag{4.32}
\end{equation}
depending on the (invariant) $I_{o}=t$ only. Then, Eqs. (1.1)--(1.2) furnish
the reduced system

\begin{equation}
U_{1t}+\frac{\stackrel{.}{g}}{g}U_{1}=0,  \tag{4.33a}
\end{equation}
\begin{equation}
U_{2t}+\frac{\stackrel{.}{h}}{h}U_{2}=0,  \tag{4.33b}
\end{equation}
\begin{equation}
U_{3t}+\frac{\stackrel{.}{r}}{r}U_{3}=0,  \tag{4.33c}
\end{equation}
\begin{equation}
\frac{\stackrel{.}{g}}{g}+\frac{\stackrel{.}{h}}{h}+\frac{\stackrel{.}{r}}{r}%
=0.  \tag{4.33d}
\end{equation}
These equations give

\begin{equation}
U_{1}=\frac{c_{1}}{g},U_{2}=\frac{c_{2}}{h},U_{3}=\frac{c_{3}}{r}, 
\tag{4.34}
\end{equation}
and

\begin{equation}
U_{1}U_{2}U_{3}=const.  \tag{4.35}
\end{equation}
Then, from (4.31) and (4.34) we infer

\begin{equation}
u_{1}=\frac{c_{1}}{g}+x\,\frac{\stackrel{.}{g}}{g},u_{2}=\frac{c_{2}}{h}+y\,%
\frac{\stackrel{.}{h}}{h},\,u_{3}=\frac{c_{3}}{s}+z\frac{\stackrel{.}{s}}{s},
\tag{4.36}
\end{equation}
while the condition (4.33d) ensures that the equation $\nabla \cdot 
\stackrel{\rightarrow }{u}=0$ is satisfied. The solutions (4.36) imply that
the vorticity $\stackrel{\rightarrow }{\omega }$ $=\nabla \wedge \stackrel{%
\rightarrow }{u}$ vanishes, so that the motion is irrotational. Thus, we may
introduce the velocity potential $\phi $ defined by $\stackrel{\rightarrow }{%
u}=\nabla \phi .$ We have

\begin{equation}
\phi =\frac{1}{2}(x^{2}\frac{\stackrel{.}{g}}{g}+y^{2}\frac{\stackrel{.}{h}}{%
h}+z^{2}\frac{\stackrel{.}{s}}{s})+\frac{c_{1}}{g}x+\frac{c_{2}}{h}y+\frac{%
c_{3}}{s}z+\phi _{o}.  \tag{4.37}
\end{equation}

\noindent Of course, $\phi $ fulfils the Laplace equation $\nabla ^{2}\phi
=0.$

\subsection{\protect\smallskip {\bf Case iv)}}

\bigskip \qquad The generator $V_{5}$ gives rise to the finite group
transformations

\[
x^{^{\prime }}=x\exp (\epsilon ),y^{^{\prime }}=y\exp (\epsilon
),z^{^{\prime }}=z\exp (\epsilon ),t^{^{\prime }}=t\exp (2\epsilon
),u_{1}^{^{\prime }}=u_{1}\exp (-\epsilon ), 
\]

\begin{equation}
u_{2}^{^{\prime }}=u_{2}\exp (-\epsilon ),\,u_{3}^{^{\prime }}=u_{3}\exp
(-\epsilon ),p^{^{\prime }}=p\exp (-2\epsilon ).  \tag{4.38}
\end{equation}

A set of basis invariants is

\[
\xi =\frac{x}{y},\eta =\frac{x}{z},\theta =\frac{x^{2}}{t},\Lambda _{1}(\xi
,\eta ,\theta )=u_{1}x,\Lambda _{2}(\xi ,\eta ,\theta )=u_{2}x, 
\]

\begin{equation}
\Lambda _{3}(\xi ,\eta ,\theta )=u_{3}x,\pi (\xi ,\eta ,\theta )=px^{2}. 
\tag{4.39}
\end{equation}

The reduced equations read

\[
-2\pi +2\theta \pi _{\theta }+\eta \pi _{\eta }+\xi \pi _{\xi }-2\nu \Lambda
_{1}-\Lambda _{1}^{2}+(2\nu -\theta )\theta \Lambda _{1\theta }+2\theta
\Lambda _{1}\Lambda _{1\theta }-4\theta ^{2}\nu \Lambda _{1\theta \theta }+ 
\]

\[
2\nu \eta (1-\eta ^{2})\Lambda _{1\eta }+\eta \Lambda _{1}\Lambda _{1\eta
}-\eta ^{2}\Lambda _{3}\Lambda _{1\eta }-4\nu \eta \theta \Lambda _{1\eta
\theta }-\nu \eta ^{2}(1+\eta ^{2})\Lambda _{1\eta \eta }+2\nu \xi (1-\xi
^{2})\Lambda _{1\xi }+ 
\]
\begin{equation}
\xi \Lambda _{1}\Lambda _{1\xi }-\xi ^{2}\Lambda _{2}\Lambda _{1\xi }-4\nu
\xi \theta \Lambda _{1\xi \theta }-2\nu \xi \eta \Lambda _{1\xi \eta }-\nu
\xi ^{2}(1+\xi ^{2})\Lambda _{1\xi \xi }=0,  \tag{4.40a}
\end{equation}

\[
-2\nu \Lambda _{2}-\Lambda _{1}\Lambda _{2}+(2\nu -\theta )\theta \Lambda
_{2\theta }+2\theta \Lambda _{1}\Lambda _{2\theta }-4\theta ^{2}\nu \Lambda
_{2\theta \theta }+ 
\]

\[
2\nu \eta (1-\eta ^{2})\Lambda _{2\eta }+\eta \Lambda _{1}\Lambda _{2\eta
}-\eta ^{2}\Lambda _{3}\Lambda _{2\eta }-4\nu \eta \theta \Lambda _{2\eta
\theta }-\nu \eta ^{2}(1+\eta ^{2})\Lambda _{2\eta \eta }+2\nu \xi (1-\xi
^{2})\Lambda _{2\xi }+ 
\]

\begin{equation}
\xi \Lambda _{1}\Lambda _{2\xi }-\xi ^{2}\Lambda _{2}\Lambda _{2\xi }-4\nu
\xi \theta \Lambda _{2\xi \theta }-2\nu \xi \eta \Lambda _{2\xi \eta }-\nu
\xi ^{2}(1+\xi ^{2})\Lambda _{2\xi \xi }-\xi ^{2}\pi _{\xi }=0,  \tag{4.40b}
\end{equation}

\[
-2\nu \Lambda _{3}-\Lambda _{1}\Lambda _{3}+(2\nu -\theta )\theta \Lambda
_{3\theta }+2\theta \Lambda _{1}\Lambda _{3\theta }-4\theta ^{2}\nu \Lambda
_{3\theta \theta }+ 
\]

\[
2\nu \eta (1-\eta ^{2})\Lambda _{3\eta }+\eta \Lambda _{1}\Lambda _{3\eta
}-\eta ^{2}\Lambda _{3}\Lambda _{3\eta }-4\nu \eta \theta \Lambda _{3\eta
\theta }-\nu \eta ^{2}(1+\eta ^{2})\Lambda _{3\eta \eta }+2\nu \xi (1-\xi
^{2})\Lambda _{3\xi }+ 
\]

\begin{equation}
\xi \Lambda _{1}\Lambda _{3\xi }-\xi ^{2}\Lambda _{2}\Lambda _{3\xi }-4\nu
\xi \theta \Lambda _{3\xi \theta }-2\nu \xi \eta \Lambda _{3\xi \eta }-\nu
\xi ^{2}(1+\xi ^{2})\Lambda _{3\xi \xi }-\eta ^{2}\pi _{\eta }=0, 
\tag{4.40c}
\end{equation}

\begin{equation}
-\Lambda _{1}+2\theta \Lambda _{1\theta }+\eta \Lambda _{1\eta }+\xi \Lambda
_{1\xi }-\eta ^{2}\Lambda _{3\eta }-\xi ^{2}\Lambda _{2\xi }=0.  \tag{4.40d}
\end{equation}

A special solution of this huge nonlinear system of partial differential
equations is derived in Appendix B. It reads

\begin{equation}
u_{1}=ct^{-\frac{1}{2}},  \tag{4.41a}
\end{equation}

\[
u_{2}=-2at^{-\frac{1}{2}}+c_{1}(\nu t^{-1})^{\frac{1}{2}}\exp [\frac{xt^{-%
\frac{1}{2}}}{4\nu }(4c-xt^{-\frac{1}{2}})]-c_{2}(\nu t^{-1})^{\frac{1}{2}%
}\exp [\frac{xt^{-\frac{1}{2}}}{4\nu }(4c-xt^{-\frac{1}{2}})]\times 
\]

\begin{equation}
\func{erf}i[\frac{1}{2}\nu ^{-\frac{1}{2}}(2c-xt^{-\frac{1}{2}})], 
\tag{4.41b}
\end{equation}

\[
u_{3}=-2bt^{-\frac{1}{2}}+c_{3}(\nu t^{-1})^{\frac{1}{2}}\exp [\frac{xt^{-%
\frac{1}{2}}}{4\nu }(4c-xt^{-\frac{1}{2}})]-c_{4}(\nu t^{-1})^{\frac{1}{2}%
}\exp [\frac{xt^{-\frac{1}{2}}}{4\nu }(4c-xt^{-\frac{1}{2}})]\times 
\]

\begin{equation}
\func{erf}i[\frac{1}{2}\nu ^{-\frac{1}{2}}(2c-xt^{-\frac{1}{2}})], 
\tag{4.41c}
\end{equation}

\begin{equation}
p=c_{o}t^{-1}+t^{-\frac{3}{2}}(\frac{c}{2}x-ay-bz),  \tag{4.41d}
\end{equation}

\noindent where

\begin{equation}
\func{erf}(\zeta )=\frac{2}{\sqrt{\pi }}\int_{0}^{\zeta }\exp (-t^{2})dt, 
\tag{4.42}
\end{equation}

\begin{equation}
\func{erf}i(\zeta )=-i\func{erf}(i\zeta )=\frac{-2i}{\sqrt{\pi }}%
\int_{0}^{i\zeta }\exp (-t^{2})dt,  \tag{4.43}
\end{equation}

\noindent and $a,b,c$ and $c_{j}$ $(j=0,1,2,3,4)$ are constants.

\bigskip The vorticity corresponding to the solution (4.41a)--(4.41d) is

\begin{equation}
\stackrel{\rightarrow }{\omega }=(0,-\frac{2}{t}\Lambda _{3\theta },-\frac{1%
}{x^{2}}\Lambda _{2}+\frac{2}{t}\Lambda _{2\theta })^{T},  \tag{4.44}
\end{equation}

\noindent where $\Lambda _{2}(\theta )$ and $\Lambda _{3}(\theta )$ are
expressed in Appendix B. We notice that in this case we obtain a stretching
rate coinciding with the null vector, i.e. $S\stackrel{\rightarrow }{\omega }%
\equiv (0,0,0)^{T},$ so that the dynamic angle $\phi =\arctan (\frac{\chi }{%
\alpha })$ is not definite.

\bigskip

\subsection{\bf Case v)}

\smallskip

Let us consider the symmetry operator $V_{6}.$ The corresponding group
transformations are

\begin{equation}
x^{^{\prime }}=x\cos \varepsilon +y\sin \varepsilon ,  \tag{4.45a}
\end{equation}

\begin{equation}
y^{^{\prime }}=-x\sin \varepsilon +y\cos \varepsilon ,  \tag{4.45b}
\end{equation}
\begin{equation}
z^{\prime }=z,  \tag{4.45c}
\end{equation}
\begin{equation}
t^{\prime }=t,  \tag{4.45d}
\end{equation}
\begin{equation}
U_{1}=u_{1}\cos \varepsilon +u_{2}\sin \varepsilon ,  \tag{4.45e}
\end{equation}
\begin{equation}
U_{2}=-u_{1}\sin \varepsilon +u_{2}\cos \varepsilon ,  \tag{4.45f}
\end{equation}

\begin{equation}
U_{3}=u_{3},  \tag{4.45g}
\end{equation}

\begin{equation}
p^{\prime }=p  \tag{4.45h}
\end{equation}

\noindent which leave the quantities

\begin{equation}
z,t,r=(x^{2}+y^{2})^{\frac{1}{2}},  \tag{4.46}
\end{equation}

\begin{equation}
\pi =p  \tag{4.47}
\end{equation}

\noindent invariant.

Putting

\begin{equation}
x=r\cos \varepsilon ,\,y=r\sin \varepsilon ,  \tag{4.48a}
\end{equation}
\begin{equation}
u_{1}=U_{1}\cos \varepsilon -U_{2}\sin \varepsilon ,  \tag{4.48b}
\end{equation}
\begin{equation}
u_{2}=U_{1}\sin \varepsilon +U_{2}\cos \varepsilon ,  \tag{4.48c}
\end{equation}
\begin{equation}
u_{3}=U_{3},  \tag{4.48d}
\end{equation}
where $U_{j}=U_{j}(r,z,t),$ Eqs. (1.1)--(1.2) take the form

\begin{equation}
U_{1t}+U_{1}U_{1r}-\frac{U_{2}^{2}}{r}+U_{1z}U_{3}+\pi _{r}-\nu (U_{1rr}+%
\frac{U_{1r}}{r}-\frac{U_{1}}{r^{2}}+U_{1zz})=0,  \tag{4.49}
\end{equation}
\begin{equation}
U_{2t}+U_{1}U_{2r}+\frac{U_{1}U_{2}}{r}+U_{2z}U_{3}-\nu (U_{2rr}+\frac{U_{2r}%
}{r}-\frac{U_{2}}{r^{2}}+U_{2zz})=0,  \tag{4.50}
\end{equation}

\begin{equation}
U_{1r}+\frac{U_{1}}{r}+U_{3z}=0,  \tag{4.51}
\end{equation}
\begin{equation}
U_{3t}+U_{1}U_{3r}+U_{3}U_{3z}-\nu (U_{3rr}+\frac{U_{3r}}{r}+U_{3zz})+\pi
_{z}=0.  \tag{4.52}
\end{equation}

\smallskip

To provide an example of exact solution to the reduced system
(4.49)--(4.52), let us look for a solution such that

\begin{equation}
U_{1}=\frac{\alpha _{o}}{r},  \tag{4.53a}
\end{equation}

\begin{equation}
U_{2}=\frac{\beta _{o}}{r},  \tag{4.53b}
\end{equation}

\noindent where $\alpha _{o},\beta _{o}$ are constants. Then, Eq. (4.49)
yields

\begin{equation}
\pi _{r}=\frac{\alpha _{o}-\beta _{o}}{r^{3}},  \tag{4.53c}
\end{equation}

\noindent while Eq. (4.50) turns out to be identically satisfied.
Furthermore, Eq. (4.51) tells us that $U_{3z}=0,$ i.e. $U_{3}=U_{3}(r,t).$
Consequently, Eq. (4.52) becomes

\begin{equation}
U_{3t}+\frac{\alpha _{o}-\nu }{r}U_{3r}-\nu U_{3rr}=-\pi _{z}.  \tag{4.54}
\end{equation}

The compatibility condition $\pi _{rz}=\pi _{zr}$ gives

\begin{equation}
\partial _{r}(U_{3t}+\frac{\alpha _{o}-\nu }{r}U_{3r}-\nu U_{3rr})=0. 
\tag{4.55}
\end{equation}

By integrating (4.53c) with respect to $r$ we get

\begin{equation}
\pi =-\frac{\alpha _{o}^{2}+\beta _{o}^{2}}{2r^{2}}+F(z,t),  \tag{4.56}
\end{equation}

\noindent $F(z,t)$ being a function of integration. Since $\pi _{z}=F_{z},$
we deduce that $F(z,t)$ has to be of the form

\begin{equation}
F(z,t)=a(t)z+b(t),  \tag{4.57}
\end{equation}

\noindent where $a,b$ are function of time. Hence

\begin{equation}
U_{3t}+\frac{\alpha _{o}-\nu }{r}U_{3r}-\nu U_{3rr}=-a_{o},  \tag{4.58}
\end{equation}

\noindent where we have supposed $a=a_{o}=const,b=b_{o}=const$. From (4.56)
we obtain the expression for the pression $p,$ which reads

\begin{equation}
p\equiv \pi =-\frac{\alpha _{o}^{2}+\beta _{o}^{2}}{2r^{2}}+a_{o}z+b_{o}. 
\tag{4.59}
\end{equation}

To find $U_{3},$ we shall distinguish two cases. Precisely:

\bigskip

{\bf I)} $\alpha _{o}=\nu .$

\bigskip

Then, Eq. (4.58) takes the form of the heat equation with a constant source,
namely

\begin{equation}
U_{3t}-\nu U_{3rr}=-a_{o}.  \tag{4.60}
\end{equation}

\bigskip A solution of Eq. (4.60) is given by

\begin{equation}
U_{3}=\frac{\gamma _{o}}{\sqrt{\nu t}}\exp (-\frac{r^{2}}{4\nu t})+\frac{%
a_{o}}{2\nu }r^{2},  \tag{4.61}
\end{equation}

\noindent where $\gamma _{o}$ is a constant. Thus, collecting all the
information, we find that the Navier--Stokes equations (1.1)--(1.2) afford
the solution

\begin{equation}
u_{1}=\frac{\nu x-\beta _{o}y}{r^{2}},  \tag{4.62a}
\end{equation}

\begin{equation}
u_{2}=\frac{\beta _{o}x+\nu y}{r^{2}},  \tag{4.62b}
\end{equation}

$u_{3}=U_{3}$ (see (4.48d)), and (see (4.61))

\begin{equation}
p=-\frac{\nu ^{2}+\beta _{o}^{2}}{2r^{2}}+a_{o}\,z+b_{o}.  \tag{4.62c}
\end{equation}

\bigskip {\bf II)} $\alpha _{o}\neq \nu .$

\bigskip In this case Eq. (4.58) can be transformed into the equation

\begin{equation}
\Psi _{t}+\frac{\alpha _{o}-\nu }{r}\Psi _{r}-\nu \Psi _{rr}=0  \tag{4.63}
\end{equation}

\noindent via the change of variable

\begin{equation}
U_{3}=\Psi (r,t)-a_{o}t.  \tag{4.64}
\end{equation}

By setting

\begin{equation}
\Psi =M(t)\,N(r),  \tag{4.65}
\end{equation}

\bigskip

\noindent Eq. (4.63) is splitted into the ordinary differential equations

\bigskip 
\begin{equation}
M_{t}=-\delta \,M,  \tag{4.66a}
\end{equation}

\begin{equation}
\nu N_{rr}-\frac{\alpha _{o}-\nu }{r}N_{r}+\delta N=0,  \tag{4.66b}
\end{equation}

\noindent where $\delta >0$ is a constant. These equations can be easily
solved to give

\begin{equation}
M=M_{o}\exp (-\delta t),  \tag{4.67a}
\end{equation}

\begin{equation}
N=r^{\mu }[c_{1}J_{\mu }(\sqrt{\frac{\delta }{\nu }}r)+c_{2}Y_{\mu }(\sqrt{%
\frac{\delta }{\nu }}r)],  \tag{4.67b}
\end{equation}

\noindent where $M_{o},c_{1},c_{2}$ are constant, $\mu =\frac{\alpha _{o}}{%
2\nu },$ and $J_{\mu },Y_{\mu }$ denote the Bessel functions of the first
and the second kind, respectively. Substituting (4.67a) and (4.67b) into
(4.64) provides

\begin{equation}
u_{3}=U_{3}=M_{o}\exp (-\delta t)r^{\mu }[c_{1}J_{\mu }(\sqrt{\frac{\delta }{%
\nu }}r)+c_{2}Y_{\mu }(\sqrt{\frac{\delta }{\nu }}r)]-a_{o}t.  \tag{4.68}
\end{equation}

The components $u_{1},u_{2}$ and the pression $p$ are

\begin{equation}
u_{1}=\frac{\alpha _{o}x-\beta _{o}y}{r^{2}},  \tag{4.69a}
\end{equation}

\begin{equation}
u_{2}=\frac{\alpha _{o}y+\beta _{o}x}{r^{2}},  \tag{4.69b}
\end{equation}

\begin{equation}
p=-\frac{\alpha _{o}^{2}+\beta _{o}^{2}}{2r^{2}}+a_{o}z+b_{o},  \tag{4.69c}
\end{equation}

\noindent respectively.

\subsubsection{The Burgers vortex and other solutions}

\bigskip

It is noteworthy that the infinitesimal\ operator V$_{6}\;$ leads to the
Burgers vortex solution [14] as a special case, namely:

\begin{equation}
u_{1}=-\frac{\gamma }{2}\,x-y\,f(r)  \tag{4.70a}
\end{equation}

\begin{equation}
u_{2}=-\frac{\gamma }{2}\,y+x\,f(r)  \tag{4.70b}
\end{equation}

\begin{equation}
u_{3}=\gamma \,z,  \tag{4.70c}
\end{equation}

\noindent where 
\begin{equation}
f(r)=f_{0}\frac{1-\exp \left( -ar^{2}\right) }{r^{2}},  \tag{4.71}
\end{equation}

\noindent $\gamma \;$and\ $f_{0}\;$are constants, and

\begin{equation}
a=\frac{\gamma }{4\,\nu }.  \tag{4.72}
\end{equation}

In fact, by choosing

\begin{equation}
U_{1}=-\frac{\gamma }{2}\,r,\,U_{2}=r\,f(r),\,U_{3}=\gamma \,z,  \tag{4.73}
\end{equation}

\noindent the reduced equations (4.49)--(4.52) are satisfied provided that

\begin{equation}
\pi _{r}=r\,f^{\,2}\left( r\right) -\frac{\gamma ^{2}}{4}r,\,\pi
_{z}=-\gamma ^{2\,}z,  \tag{4.74}
\end{equation}

\noindent where $\,f(r)\;$obeys the equation 
\begin{equation}
f_{rr}+\left( \frac{\gamma }{2\,\nu }r+\frac{3}{r}\right) f_{r}+\frac{\gamma 
}{\nu }\,f=0.  \tag{4.75}
\end{equation}

The general solution of Eq (4.75) can be written as

\begin{equation}
f\left( r\right) =\frac{f_{0}-\left( f_{0}-f_{1}\right) \exp \left(
-ar^{2}\right) }{r^{2}},  \tag{4.76}
\end{equation}

\noindent which reproduces just (4.71) for $f_{1}=0.\;$Eqs. (4.70a)--(4.70c)
are derived from (4.49)--(4.52), with the help of (4.73).\ 

Eqs. (4.74) give the pression $(p=\pi ):$

\begin{equation}
p=-\frac{\gamma ^{2}}{2}\left( z^{2}+\frac{r^{2}}{4}\right) -\frac{%
f_{\,0}^{\,2}}{2\,r^{2}}\left[ 1-\exp \left( -ar^{2}\right) \right]
^{2}+af\,_{\,0}^{2}\left[ \func{Ei}\left( -ar^{2}\right) -\func{Ei}\left(
-2ar^{2}\right) \right] ,  \tag{4.77}
\end{equation}

\noindent where $\func{Ei}\;$denotes the exponential--integral function [11]

\[
\func{Ei}\left( \zeta \right) =-\int_{-\zeta }^{\infty }\frac{\exp \left(
-t\right) }{t}dt, 
\]

\noindent while the vorticity is 
\begin{equation}
\overrightarrow{\omega }=\left( 0,0,\omega _{3}\right) ^{T}  \tag{4.78}
\end{equation}

\noindent with 
\begin{equation}
\omega _{3}=2\,f+r\,f_{r}=2\,a\,f_{0}\,\exp \left( -a\,r^{2}\right) . 
\tag{
4.79}
\end{equation}

We point out that (4.79) coincides with the expression for the Burgers
vortex [14], in which there exists a balance of dissipation and stretching
[3], [15], [4].

Now by integrating Eqs. (4.70a) and (4.70c) we easily find 
\begin{equation}
r=r_{0}\exp \left( -\frac{\gamma }{2}\,t\right) ,  \tag{4.80}
\end{equation}
which tells us that in the limit $t\rightarrow \infty ,\,r\rightarrow 0.\;$%
For $r\rightarrow 0\;$both $f\left( r\right) \;$given by (4.71) and the
vorticity (4.79) turn out to be finite.

\subsubsection{\protect\bigskip The Burgers-Lundgren solution}

\bigskip

Another interesting solution of the Navier-Stokes equations (1.1)--(1.2) of
the Burgers vortex type, generated by the vector field V$_{6},\;$is obtained
assuming that 
\begin{equation}
U_{1}=-\frac{\gamma }{2}\,r,\,U_{2}=r\,f(r,t),\,U_{3}=\gamma \,z,  \tag{4.81}
\end{equation}
where the function $f(r,t)\;$depends on the time also.

In doing so, substitution from (4.139) into Eqs. (4.49)-(4.52) gets

\begin{equation}
f_{t}-\nu \,f_{rr}-\left( \frac{\gamma }{2}r+\frac{3\nu }{r}\right)
f_{r}-\gamma \,f=0.  \tag{4.82}
\end{equation}

By inspection, Eq. (4.82) is satisfied by

\begin{equation}
f\,\left( r,t\right) =f_{0}\frac{1-\exp \left[ -ar^{2}\left( \frac{1}{1-\exp
\left( -\gamma \,t\right) }\right) \right] }{r^{2}}.  \tag{4.83}
\end{equation}

In this case, the vorticity is time dependent and takes the form 
\[
\overrightarrow{\omega }=\left( 0,0,\omega _{3}\left( r,t\right) \right)
^{T}, 
\]

\noindent with

\begin{equation}
\omega _{3}\left( r,t\right) =2f\left( r,t\right) +rf_{r}\left( r,t\right)
=2\,a\,f_{0}\frac{\exp \left( 1-\frac{ar^{2}}{1-\exp \left( -\gamma
\,t\right) }\right) }{1-\exp \left( -\gamma \,t\right) }.  \tag{4.84}
\end{equation}

We remark that this quantity is the same as the expression of the vorticity
given by Lundgren [15, formula (16)]. When $t\rightarrow \infty ,\,\omega
_{3}\left( r,t\right) \;$tends to the Burgers vortex (4.71). Therefore, as
it was noted in [15] one can resort to the assumption of Townsend [16], who
considered the axial strain rate proportional to the root-mean-square strain
rate in turbulent flow. Under this hypotesis, the Burgers vortex leads
naturally to the Kolmogorov lenght 
\[
\eta =\left( \frac{\nu ^{3}}{\epsilon }\right) ^{1/4}, 
\]

\noindent where $\epsilon \;$is the dissipation rate per unit mass.

\subsubsection{Another vortex-like solution of Eq. (4.82)}

Let us look for a solution of Eq. (4.82) of the form 
\begin{equation}
f=A\left( t\right) \exp [B\left( t\right) r^{2}],  \tag{4.85}
\end{equation}

\noindent where $A$ and $B$ are functions of time to be determined.
Inserting (4.85) into Eq. (4.82) yields 
\[
A=\frac{A_{0}}{[\exp \frac{\gamma }{2}t-\frac{c_{1}}{c_{0}\gamma }\exp
\left( -\frac{\gamma }{2}t\right) ]^{2}}, 
\]
\[
B=-\frac{a}{[1-\frac{c_{1}}{c_{0}\gamma }\exp \left( -\frac{\gamma }{2}%
t\right) ]}, 
\]

\noindent with $a=\frac{\gamma }{4\nu }.$

By choosing $\frac{c_{1}}{c_{0}\gamma }=1,$we obtain 
\begin{equation}
f\left( r,t\right) =\frac{f_{0}}{2}\sec h^{2}\left( \frac{\gamma }{2}%
t\right) \exp \left( -\frac{a\,r^{2}}{1+\exp \left( -\gamma \,t\right) }%
\right) .  \tag{4.86}
\end{equation}

In opposition to the behavior of (4.83) and (4.84), this solution tends to
zero when $t\rightarrow \infty .$

\bigskip

\subsubsection{Evaluation of the dynamic angles for vortices related to $%
V_{6}$}

\bigskip

$1)$ The solution (4.61)-(4.62b) has a vorticity given by $\stackrel{%
\rightarrow }{\omega }=(\omega _{1},\omega _{2},0)^{T},$ with

\begin{equation}
\omega _{1}=[\frac{a_{o}}{\nu }-\frac{\gamma _{o}}{2}(\nu t)^{-\frac{3}{2}%
}\exp (-\frac{r^{2}}{4\nu t})]y,\omega _{2}=[-\frac{a_{o}}{\nu }+\frac{%
\gamma _{o}}{2}(\nu t)^{-\frac{3}{2}}\exp (-\frac{r^{2}}{4\nu t})]x, 
\tag{4.87}
\end{equation}

\noindent which leads to a constant dynamic angle, namely

\begin{equation}
\phi =\arctan \frac{\beta _{o}}{\nu }.  \tag{4.88}
\end{equation}

The vorticity lies on the $x,y-$plane and is ortogonal to $\stackrel{%
\rightarrow }{r}=x\,\widehat{x}+y\,\widehat{y},$ i.e. $\stackrel{\rightarrow 
}{\omega }\cdot \stackrel{\rightarrow }{r}=0.$ The orientation of $\stackrel{%
\rightarrow }{\omega }$ \ with respect to $S\stackrel{\rightarrow }{\omega }$
depends on the ratio $\frac{\beta _{o}}{\nu };$ for $\beta _{o}=0,$ $\phi
=0; $ for $\beta _{o}=\nu ,$ $\phi =\frac{\pi }{4},$ while in absence of
viscosity ($\nu \rightarrow 0)$ and when $\beta _{o}\neq 0$ we have $\phi =%
\frac{\pi }{2}.$

\bigskip

2) Let us consider the solution (4.68)-(4.69b). Owing to the complicated
structure of $u_{3}$ (see (4.68)), the expression of the vorticity results
too cumbersome. Therefore, it is convenient to examine the asymptotic
behaviour of $u_{3}$ for $r\rightarrow \infty $ at fixed $\mu .$ By choosing 
$a_{o}=0,$ we get [11, p. 364]

\begin{equation}
u_{3}\sim \exp (-\delta t)r^{\mu -\frac{1}{2}}[c_{1}\cos (\sqrt{\frac{\delta 
}{\nu }}r-\mu \frac{\pi }{2}-\frac{\pi }{4})+c_{2}\sin (\sqrt{\frac{\delta }{%
\nu }}r-\mu \frac{\pi }{2}-\frac{\pi }{4})].  \tag{4.89}
\end{equation}

\bigskip Now the vorticity corresponding to the solution (4.69a), (4.69b)
and (4.89) and the dynamic angle can be easily calculated. For simplicity,
we report only the value of $\phi ,$ which is

\[
\phi =\arctan \frac{\beta _{o}}{\alpha _{o}}. 
\]

Thus, the discussion on this result is similar to that performed in the case
1). We observe that for $\alpha _{o}=\beta _{o}=0,$ $\phi $ is not defined.

An interesting case is represented by the choice $\mu =\frac{1}{2}$ in Eq.
(4.89). In fact, we get

\begin{equation}
u_{3}\sim \exp (-\frac{\Omega ^{2}}{2}\tau ^{2})(c_{1}\sin \Omega
t-c_{2}\cos \Omega t),  \tag{4.90}
\end{equation}

\noindent where $\Omega ^{2}=2\delta $ and $\tau =t^{\frac{1}{2}}.$
Formally, the quantity (4.90) can be interpreted as a special case of the
general solution of the parametric oscillator described by the equation

\begin{equation}
\stackrel{..}{q}+4\gamma _{1}t\stackrel{.}{q}+(\gamma _{2}^{2}+2\gamma
_{1}+4\gamma _{1}^{2}t^{2})q=0,  \tag{4.91}
\end{equation}

\noindent whose general solution is

\begin{equation}
q=\exp (-\gamma _{1}t^{2})(c_{1}\cos \gamma _{2}t+c_{2}\sin \gamma _{2}t), 
\tag{4.92}
\end{equation}

\noindent where the amplitude plays the role of a damping function ($\gamma
_{1}>0)$ which tends to zero for large values of time according to a law of
the Gaussian type.

\bigskip

3) The dynamic angles for the Burgers and Burgers-Lundgren vortices,
together with the vortex solution with $f(r,t)$ given by (4.86), are zero.

\bigskip

\section{ Invariance of the boundary conditions}

\setcounter{equation}{0}

The study of the invariance properties of a system of differential equations
with suitable boundary conditions plays a fundamental role in the
description of realistic models.

Generally, boundary value problems can be treated more simply for ordinary
differential equations. Indeed, a symmetry analysis allows one to reduce a
given boundary value problem for an ordinary differential equation to a
boundary value problem for a (reduced) equation of lower order. In the case
of a partial differential equation, by a group point of view, a boundary
value problem is solved when one can determine a solution which is
invariant, together with all the boundary conditions, under the action of
the infinitesiaml generator of the symmetry group. However, for a linear
partial differential equation one can resort to less restrictive conditions
[8, p. 215].

In this Section we present a systematic investigation of all the surfaces
which can take the meaning of invariant boundary surfaces in a boundary
value problem for the Navier-Stokes (1.1)--(1.2).

To this aim, we recall that if $V$ denotes a generator of symmetry
transformations, a basis of invariants for $V$ is obtained from the equation

\begin{equation}
VI=0.
\end{equation}

\QTP{Body Math}
Now let $V=\stackrel{9}{\stackunder{i=1}{\sum }}V_{i}$ be the vector field
generating the set of the Lie--point transformations of Eqs. (1.1)--(1.2).
Using the method of characteristics, Eq. (5.1) can be written in the form

\QTP{Body Math}
\begin{eqnarray}
\frac{dx}{ax+by+ez+g_{o}} &=&\frac{dy}{ay-bx+cz+h_{o}}=\frac{dz}{%
az-cy-ex+r_{o}}=  \nonumber \\
\frac{dt}{2at+d} &=&\frac{du_{1}}{-au_{1}+bu_{2}+eu_{3}}=\frac{du_{2}}{%
-au_{2}-bu_{1}+cu_{3}}=  \nonumber \\
\frac{du_{3}}{-au_{3}-cu_{2}-eu_{1}} &=&\frac{dp}{k_{o}-2ap}.
\end{eqnarray}

\QTP{Body Math}
Here we have put $\;g=g_{o}=const,\;h=h_{o}=const,\;k=k_{o}=const.\;$By
solving the system (5.2), we can determine the characteristic manifolds $%
\lambda =const,\;\chi =const,\;\psi =const$ in the coordinate space $%
(x,y,z,t).$ To this aim, first let us analyse the general case.
Subsequently, some special interesting case will be examined.

\QTP{Body Math}
$\bigskip $

\subsection{The general case}

The system (5.2) can be put into the form:

\begin{equation}
\frac{d\overrightarrow{x}}{d\tau }=A\cdot \overrightarrow{x}+\overrightarrow{%
r}
\end{equation}

\noindent where:

\begin{equation}
A=\left( 
\begin{array}{lll}
a & b & e \\ 
-b & a & c \\ 
-e & -c & a
\end{array}
\right)
\end{equation}

\begin{equation}
\overrightarrow{x}=\left( 
\begin{array}{l}
x \\ 
y \\ 
z
\end{array}
\right) ,\qquad \qquad \overrightarrow{r}=\left( 
\begin{array}{l}
g_{0} \\ 
h_{0} \\ 
r_{0}
\end{array}
\right)
\end{equation}

\begin{equation}
\tau =\frac{1}{2a}\ln \left( \frac{2at}{d}+1\right) .
\end{equation}

\bigskip

Let us distinguish the following cases.

\bigskip

{\bf Case I) }$\qquad a\neq 0.$

\bigskip

Eq. (5.3) admits the solution

\begin{equation}
\overrightarrow{x}=e^{A\tau }\overrightarrow{\,\,x_{0}}+\overrightarrow{%
x_{P,}}
\end{equation}

\noindent where $\overrightarrow{\,\,x_{0}}\;$is a constant vector
determined by initial conditions, and

\begin{equation}
\overrightarrow{x_{P}}=-A^{-1}\overrightarrow{r}
\end{equation}

First, let us consider the subcase $\overrightarrow{r}=0.$

The solution of the system (5.3) (in the coordinate space), is provided by
the following invariants:

\begin{eqnarray}
\lambda &=&\frac{1}{\sqrt{2a\frac{t}{d}+1}}\{(\frac{1}{\omega ^{2}}[%
c^{2}+(b^{2}+e^{2})\cos \omega \tau ])x+  \nonumber \\
&&(\frac{ce}{\omega ^{2}}[-1+\cos \omega \tau ]-\frac{b}{\omega }\sin \omega
\tau )y+  \nonumber \\
&&(-\frac{bc}{\omega ^{2}}[-1+\cos \omega \tau ]-\frac{e}{\omega }\sin
\omega \tau )z\},
\end{eqnarray}

\bigskip

\begin{eqnarray}
\mu &=&\frac{1}{\sqrt{2a\frac{t}{d}+1}}\{(\frac{ce}{\omega ^{2}}[-1+\cos
\omega \tau ]+\frac{b}{\omega }\sin \omega \tau )x+(\frac{1}{\omega ^{2}}%
[c^{2}+  \nonumber \\
&&(b^{2}+e^{2})\cos \omega \tau ])y+(\frac{be}{\omega ^{2}}[-1+\cos \omega
\tau ]-\frac{c}{\omega }\sin \omega \tau )z],
\end{eqnarray}

\begin{eqnarray}
\nu &=&\frac{1}{\sqrt{2a\frac{t}{d}+1}}\{(-\frac{bc}{\omega ^{2}}[-1+\cos
\omega \tau ]+\frac{e}{\omega }\sin \omega \tau )x+  \nonumber \\
&&(\frac{be}{\omega ^{2}}[-1+\cos \omega \tau ]+\frac{c}{\omega }\sin \omega
\tau )y+  \nonumber \\
&&(\frac{1}{\omega ^{2}}[b^{2}+(c^{2}+e^{2})\cos \omega \tau ])z\},
\end{eqnarray}

\bigskip

\noindent where $\omega =\sqrt{b^{2}+c^{2}+e^{2}}.$

From general formulas (5.9)--(5.11) we can obtain the following invariants,
in a simple form: 
\begin{equation}
\lambda =\sqrt{\frac{d}{2at+d}}(cx-ey+bz)
\end{equation}

\begin{equation}
\mu =\sqrt{\frac{d}{2at+d}}%
(ex+cy)^{2}+b^{2}(x^{2}+y^{2})+(c^{2}+e^{2})z^{2}+b(-2cxz+2eyz)
\end{equation}

\begin{equation}
\nu =\frac{(ex+cy)^{2}+b^{2}(x^{2}+y^{2})+(c^{2}+e^{2})z^{2}+b(-2cxz+2eyz)}{%
(cx-ey+bz)^{2}}
\end{equation}

The case where $g_{o}\neq 0,\;h_{o}\neq 0,\;r_{o}\neq 0$ can be obtained
from the previous one by making the change of variables:

\begin{equation}
x\rightarrow x+\frac{a^{2}g_{o}+c(cg_{o}-eh_{o}+br_{o})-a(bh_{o}+er_{o})}{%
a(a^{2}+b^{2}+c^{2}+e^{2})},
\end{equation}

\begin{equation}
y\rightarrow y+\frac{abg_{o}-ceg_{o}+(a^{2}+e^{2})h_{o}-(ac+be)r_{o}}{%
a(a^{2}+b^{2}+c^{2}+e^{2})},
\end{equation}

\begin{equation}
z\rightarrow z+\frac{bcg_{o}+aeg_{o}+(ac-be)h_{o}+(a^{2}+b^{2})r_{o}}{%
a(a^{2}+b^{2}+c^{2}+e^{2})}.
\end{equation}

\bigskip

\bigskip

\bigskip {\bf Case II) \ \ }$a=0.$

\bigskip

For $\overrightarrow{r}=0\;$ we obtain the following invariants:\bigskip 
\begin{equation}
\lambda =(cx-ey+bz)
\end{equation}
\begin{equation}
\mu =x^{2}+y^{2}+z^{2}
\end{equation}

\begin{equation}
\sqrt{b^{2}+c^{2}+e^{2}}\frac{t}{d}-\pi \arctan \frac{ex+cy}{%
b(-cx+ey)+(c^{2}+e^{2})z}
\end{equation}

Notice that $\lambda $ represents an invariant plane, whereas $\mu $ is an
invariant spheric surphace centered at the origin.

For $\;g_{o}\neq 0,\;h_{o}\neq 0,\;r_{o}\neq 0,$ we obtain:

\bigskip 
\begin{equation}
\lambda =cx-ey+bz-\frac{t}{d}(cg_{0}-eh_{0}+br_{0})
\end{equation}

\begin{eqnarray}
\mu &=&\frac{1}{\sqrt{c^{2}+e^{2}}}\cos (\omega \frac{t}{d})(-\frac{bc}{%
\omega ^{2}}g_{0}+\frac{be}{\omega ^{2}}h_{0}+\frac{c^{2}+e^{2}}{\omega ^{2}}%
r_{0}-ex-cy)+  \nonumber \\
&&\frac{1}{\sqrt{c^{2}+e^{2}}}\sin \omega \frac{t}{d}(\frac{e}{\omega }g_{0}+%
\frac{c}{\omega }h_{0}-\frac{bc}{\omega }x+\frac{be}{\omega }y+\frac{%
c^{2}+e^{2}}{\omega }z)
\end{eqnarray}

\begin{equation}
\nu =(\frac{e}{\chi }x+\frac{c}{\chi }y-\frac{bcg_{0}-beh_{0}-\chi ^{2}r_{0}%
}{\omega ^{2}\chi })^{2}+(-\frac{bc}{\omega \chi }x+\frac{be}{\omega \chi }y+%
\frac{\chi }{\omega }z+\frac{eg_{0}+ch_{0}}{\omega \chi })^{2}.
\end{equation}

\bigskip

Eq. (5.23) represents a cilinder with radius 
\[
r=\left| \lambda \right| 
\]
and axis parallel to vector

\begin{equation}
\frac{c}{\omega }\widehat{\,x}-\frac{e}{\omega }\widehat{\,y}+\frac{b}{%
\omega }\,\widehat{z}
\end{equation}
passing throw the point of coordinates:

\begin{eqnarray}
&&\frac{2b\,c\,e\,g_{0}+b\left( c^{2}-e^{2}\right) h_{0}-e\,\chi ^{2}r_{0}}{%
\omega ^{2}\chi ^{2}},\frac{g_{0}b\left( c^{2}-e^{2}\right)
-2b\,c\,e\,h_{0}-cr_{0}\chi ^{2}}{\omega ^{2}\chi ^{2}},  \nonumber \\
&&-\frac{eg_{0}+ch_{0}}{\omega ^{2}}
\end{eqnarray}

In the space of velocities, the invariant curves, which are now functions of 
$u_{1},u_{2},u_{3}$ and $p,$can be deduced by the previous ones by
performing the substitution

\begin{equation}
a\rightarrow -a,\,x\rightarrow u_{1},\,y\rightarrow u_{2},\,z\rightarrow
u_{3},\,t\rightarrow p,\,d\rightarrow k_{o}.
\end{equation}

At this point let us examine some particular cases which emerge by demanding
that some arbitrary constants appearing in (5.2) are vanishing. Precisely,
we shall distinguish the cases:

\bigskip

{\bf Subcase I}: $\;\xi _{4}\neq 0,$ where

I$1$ $:$ $d\neq 0,\;a=b=c=e=0,$

I$2:$ $d\neq 0,\;a=b=e=0,\;c\neq 0,$

I$3:$ $d\neq 0,\;a=b=0,\;c\neq 0,\;e\neq 0,$

I$4:$ $d\neq 0,\;a=0,\;b\neq 0,\;c\neq 0,\;e\neq 0.$

\bigskip

{\bf Subcase II}: $\;\xi _{4}=0,$ with\ $a=d=0,\;$and

II$1:$ $b=e=c=0,$

II$2:$ $b\neq 0,\;e=c=0,$

II$3:$ $b\neq 0,\;c\neq 0,\;e=0,$

II$4:$ $b\neq 0,\;c\neq 0,\;e\neq 0.$

\bigskip

The expressions for the invariants $\lambda ,\;\chi $ and $\psi $
corresponding to the subcases I1,...,I4 and II1,...,II4 are given by

\bigskip

{\it Subcase I1}

\bigskip 
\begin{equation}
\lambda =t-\frac{d}{g_{o}}x,
\end{equation}
\begin{equation}
\chi =y-\frac{h_{o}}{g_{o}}x,
\end{equation}
\begin{equation}
\psi =z-\frac{r_{o}}{g_{o}}x.
\end{equation}

\bigskip

{\it Subcase I2}

\bigskip 
\begin{equation}
\lambda =t-\frac{d}{g_{o}}x,
\end{equation}

\begin{equation}
\chi =\frac{h_{o}+cz}{c}\cos \frac{cx}{g_{o}}+\frac{cy-r_{o}}{c}\sin \frac{cx%
}{g_{o}},
\end{equation}

\begin{equation}
\psi =\frac{cy-r_{o}}{c}\cos \frac{cx}{g_{o}}-\frac{h_{o}+cz}{c}\sin \frac{cx%
}{g_{o}}.
\end{equation}

\bigskip

{\it Subcase I3}

\bigskip

\begin{equation}
\lambda =-\frac{c}{e}x+y,
\end{equation}

\begin{equation}
\chi =x^{2}+y^{2}+z^{2},
\end{equation}

\begin{equation}
\psi =t+\frac{d}{\sqrt{c^{2}+e^{2}}}\arctan \frac{z\sqrt{c^{2}+e^{2}}}{ex+cy}%
.
\end{equation}

\bigskip

{\it Subcase I4}

\bigskip

\begin{equation}
\lambda =\frac{cx-ey}{b}+z,
\end{equation}

\begin{equation}
\chi =x^{2}+y^{2}+z^{2},
\end{equation}

\begin{equation}
\psi =t-\frac{d}{\sqrt{b^{2}+c^{2}+e^{2}}}\arctan \frac{b^{2}x+e(ex+cy)-bcz}{%
\sqrt{b^{2}+c^{2}+e^{2}}(by+ez)}.
\end{equation}

\bigskip {\it Subcase II1}

\begin{equation}
\lambda =t,
\end{equation}

\begin{equation}
\chi =y-\frac{h_{o}}{g_{o}}x,
\end{equation}

\begin{equation}
\psi =z-\frac{r_{o}}{g_{o}}x.
\end{equation}

{\it Subcase II2}

\bigskip

\begin{equation}
\lambda =t,
\end{equation}

\begin{equation}
\chi =(x-\frac{h_{o}}{b})\cos \frac{b}{r_{o}}z-(y+\frac{g_{o}}{b})\sin \frac{%
b}{r_{o}}z,
\end{equation}

\begin{equation}
\psi =(x-\frac{h_{o}}{b})\sin \frac{b}{r_{o}}z+(y+\frac{g_{o}}{b})\cos \frac{%
b}{r_{o}}z.
\end{equation}

{\it Subcase II3}

\bigskip

\begin{equation}
\lambda =t,
\end{equation}

\begin{equation}
\chi =z+\frac{c}{b}x,
\end{equation}

\begin{equation}
\psi =x^{2}+y^{2}+z^{2}.
\end{equation}

\bigskip

{\it Subcase II4}

\bigskip

\begin{equation}
\lambda =t,
\end{equation}

\begin{equation}
\chi =cx-ey+bz,
\end{equation}

\begin{equation}
\psi =x^{2}+y^{2}+z^{2}.
\end{equation}

\section{\protect\bigskip Conclusions}

We summarize the main results achieved in this work. Using a
group-theoretical framework, we have established a connection between the
symmetry generators of the incompressible 3D Navier-Stokes equations and the
existence of solutions having a vortical dynamics. Precisely, by means of a
procedure of symmetry reduction of Eqs. (1.1)--(1.2), we have found some
novel (at the best of our knowledge) exact solutions which contain as
special cases important solutions of the vortex-type well-known in the
literature, as the Burgers vortex and shear-layer solutions. Such solutions
have been discussed by the point of view of their possible physical meaning.

It is suitable to recall that an approach of the algebraic and
group-theoretical type to the equations for the fluid dynamics, especially
to the incompressible Navier-Stokes equations, is not new and has been
carried out by many authors (see, for example, [10], and [17]--[22]). In
particular, the case of two space variables have been investigate more
frequently and in more detailed manner. This is due to the fact that in 2D
dimensions the possibility of defining a stream function [17] simplifies the
mathematical treatment of the problem. In this spirit, recently Ludlow,
Clarkson and Bassom [17] have obtained non-classical symmetries for the 2D
Navier-Stokes equations applying the so-called direct method [23]. On the
other hand, in [20] Fushchich, Shtelen and Slavutsky have determined
solutions of the 3D Navier-Stokes equations performing a systematic study of
the inequivalent ansaetze of codimension 1 by means of which a direct
reduction of the equations under consideration to ordinary differential
equations is obtained.

In our paper, together with the finding of exact solutions linked to the
reduction method and together with the analysis of the role that the
symmetry infinitesimal operators plays in understanding the vortical
phenomena, we have classified in the context of the group theory all the
invariant surfaces related to boundary conditions of physical interest. This
kind of research is of special utility in possible applications to problems
where the temperature is present. In particular, we mention the case in
which to the Eqs. (1.1)--(1.2) one adds a transport equation describing the
behavior of the temperature. These equations are known as the
Navier--Stokes--Fourier equations, where the real coupling among them occurs
just via the boundary conditions [24].

To conclude, we notice that in the symmetry algebra commutators associated
with the 3D Navier-Stokes Eqs. (1.1)--(1.2) arbitrary functions of
integration appear. This indicates that the algebra is infinite-dimensional.
Hence, it should been interesting to explore deeply this algebra and to
classify, for example, its one-dimensional subalgebras in order to discover
other possible solutions of the 3D Navier--Stokes equations.

\section*{Appendix A: the symmetry algebra}

\setcounter{equation}{0} \renewcommand{\thesection}{\Alph{section}} %
\addtocounter{section}{-5}

\setcounter{equation}{0}

The appearance of arbitrary functions in the vector fields (2.7)--(2.15)
ensures that the symmetry algebra of the Navier-Stokes equations is
infinite-dimensional. This algebra is defined by the commutation relations

\smallskip 
\begin{equation}
\lbrack V_{1}(g_{1}),V_{1}(g_{2})]=V_{4}(g_{2}\stackrel{..}{g_{1}}-g_{1}%
\stackrel{..}{g_{2}}),\text{ }[V_{2}(h_{1}),V_{2}(h_{2})]=V_{4}(h_{2}%
\stackrel{..}{h_{1}}-h_{1}\stackrel{..}{h_{2}}),
\end{equation}

\begin{equation}
\lbrack V_{3}(r_{1}),V_{3}(r_{2})]=V_{4}(r_{2}\stackrel{..}{r_{1}}-r_{1}%
\stackrel{..}{r_{2}}),\ \qquad [V_{4}(k_{1}),V_{4}(k_{2})]=0,
\end{equation}

\begin{equation}
\lbrack V_{1}(g),V_{2}(h)]=0,\qquad [V_{1}(g),V_{3}(r)]=0,\qquad
\end{equation}

\begin{equation}
\lbrack V_{1}(g),V_{4}(k)]=0,\qquad [V_{1}(g),V_{5}(a)]=V_{1}(ag-2a\stackrel{%
.}{g}t),
\end{equation}
\begin{equation}
\lbrack V_{1}(g),V_{6}(b)]=-V_{2}(bg),\qquad [V_{1}(g),V_{7}(c)]=0,
\end{equation}
\begin{equation}
\lbrack V_{1}(g),V_{8}(d)]=-V_{3}(dg),\qquad [V_{1}(g),V_{9}(e)]=-V_{1}(e%
\stackrel{.}{g}),
\end{equation}
\begin{equation}
\lbrack V_{2}(h),V_{3}(r)]=0,\qquad [V_{2}(h),V_{4}(k)]=0,
\end{equation}
\begin{equation}
\lbrack V_{2}(h),V_{5}(a)]=V_{2}(ah-2a\stackrel{.}{h}t),\qquad
[V_{2}(h),V_{6}(b)]=V_{1}(bh),
\end{equation}
\begin{equation}
V_{2}(h),V_{7}(c)]=-V_{3}(ch),\qquad [V_{2}(h),V_{8}(d)]=0,
\end{equation}
\begin{equation}
\lbrack V_{2}(h),V_{9}(e)]=-V_{2}(e\stackrel{.}{h}),\qquad
[V_{3}(r),V_{4}(k)]=0,
\end{equation}
\begin{equation}
\lbrack V_{3}(r),V_{5}(a)]=V_{3}(ar-2a\stackrel{.}{r}t),\qquad
[V_{3}(r),V_{6}(b)]=0,
\end{equation}
\begin{equation}
\lbrack V_{3}(r),V_{7}(c)]=V_{2}(cr),\qquad [V_{3}(r),V_{8}(d)]=V_{1}(dr),
\end{equation}
\begin{equation}
\lbrack V_{3}(r),V_{9}(e)]=-V_{3}(e\stackrel{.}{r}),\qquad
[V_{4}(k),V_{5}(a)]=-2V_{4}(ak+a\stackrel{.}{k}t),
\end{equation}
\begin{equation}
\lbrack V_{4}(k),V_{6}(b)]=0,\qquad [V_{4}(k),V_{7}(c)]=0,
\end{equation}
\begin{equation}
\lbrack V_{4}(k),V_{8}(d)]=0,\qquad [V_{4}(k),V_{9}(e)]=-V_{4}(e\stackrel{.}{%
k}),
\end{equation}
\begin{equation}
\lbrack V_{5}(a),V_{6}(b)]=0,\qquad [V_{5}(a),V_{7}(c)]=0,
\end{equation}
\begin{equation}
\lbrack V_{5}(a),V_{8}(d)]=0,\qquad [V_{5}(a),V_{9}(e)]=-2V_{9}(ae),
\end{equation}
\begin{equation}
\lbrack V_{6}(b),V_{7}(c)]=-V_{8}(bc),\qquad [V_{6}(b),V_{8}(d)]=V_{7}(bd),
\end{equation}
\begin{equation}
\lbrack V_{6}(b),V_{9}(e)]=0,\qquad [V_{7}(c),V_{8}(d)]=-V_{6}(cd),
\end{equation}
\begin{equation}
\lbrack V_{7}(c),V_{9}(e)]=0,\qquad [V_{8}(d),V_{9}(e)]=0.
\end{equation}

\smallskip

Scrutinizing these commutation rules, it emerges that they contain the
subalgebras 
\begin{equation}
\begin{array}{l}
\{V_{1,}V_{2},V_{5}\},\{V_{1,}V_{2},V_{6}\},\{V_{1,}V_{2},V_{9}\},%
\{V_{1,}V_{3},V_{5}\},\{V_{1,}V_{3},V_{8}\},\{V_{1,}V_{3},V_{9}\}, \\ 
\{V_{1,}V_{4},V_{5}\},\{V_{1,}V_{5},V_{9}\},\{V_{2,}V_{3},V_{5}\},%
\{V_{2,}V_{3},V_{7}\},\{V_{2,}V_{3},V_{9}\},\{V_{2,}V_{4},V_{5}\}, \\ 
\{V_{2,}V_{4},V_{9}\},\{V_{2,}V_{5},V_{9}\},\{V_{3,}V_{4},V_{5}\},%
\{V_{3,}V_{4},V_{9}\},\{V_{3,}V_{5},V_{9}\},\{V_{4,}V_{5},V_{9}\}, \\ 
\{V_{6,}V_{7},V_{8}\},
\end{array}
\end{equation}

\noindent which turn out to be isomorphic to the algebra of the Euclidean
group $E_{2}$.

\section*{Appendix B: derivation of the solution (4.41a)--(4.41d)}

\setcounter{equation}{0} \renewcommand{\thesection}{\Alph{section}} %
\addtocounter{section}{+1}

\setcounter{equation}{0}

\bigskip In order to find a particular solution of Eqs. (4.41a)--(4.41d),
let us assume that

\begin{equation}
\Lambda _{1}=h(\theta ),\Lambda _{2}=f(\theta ),\Lambda _{3}=g(\theta ).
\end{equation}

Then, Eq. (4.41d) provides

\begin{equation}
\Lambda _{1}=c\theta ^{\frac{1}{2}},
\end{equation}

\noindent $c$ being a constant of integration, while Eqs. (4.41a)--(4.41c)
give

\begin{equation}
2\pi -2\theta \pi _{\theta }-\eta \pi _{\eta }-\xi \pi _{\xi }+\frac{1}{2}%
c\theta ^{\frac{3}{2}}=0,
\end{equation}

\begin{equation}
\pi _{\xi }=\frac{1}{\xi ^{2}}F(\theta ),
\end{equation}

\begin{equation}
\pi _{\eta }=\frac{1}{\eta ^{2}}G(\theta ),
\end{equation}

\noindent respectively, where

\begin{equation}
F(\theta )=-(2\nu +c\theta ^{\frac{1}{2}})f+(2\nu \theta +2c\theta ^{\frac{3%
}{2}}-\theta ^{2})f^{\prime }-4\nu \theta ^{2}f^{\prime \prime },
\end{equation}

\begin{equation}
G(\theta )=-(2\nu +c\theta ^{\frac{1}{2}})g+(2\nu \theta +2c\theta ^{\frac{3%
}{2}}-\theta ^{2})g^{\prime }-4\nu \theta ^{2}g^{\prime \prime },
\end{equation}

\noindent with $f^{\prime }=\frac{df}{d\theta },g^{\prime }=\frac{dg}{%
d\theta }.$

By integrating Eq. (B.4) with respect to $\xi ,$ we get

\begin{equation}
\pi =-\frac{1}{\xi }F(\theta )+\alpha (\eta ,\theta ),
\end{equation}

\noindent where $\alpha (\eta ,\theta )$ is a function of integration.

From (B.8) and (B.5) we have

\begin{equation}
\pi _{\eta }=\alpha _{\eta }=\frac{G(\theta )}{\eta ^{2}},
\end{equation}

\noindent from which

\begin{equation}
\alpha =-\frac{G(\theta )}{\eta }+\beta (\theta ),
\end{equation}

\noindent where $\beta $ is an arbitrary function of $\theta .$ Taking
account of (B.10), Eq. (B.8) becomes

\begin{equation}
\pi =-\frac{F(\theta )}{\xi }-\frac{G(\theta )}{\eta }+\beta (\theta ).
\end{equation}

Then, substitution from (B.11) into Eq. (B.3) yields

\begin{equation}
F=a\theta ^{\frac{3}{2}},
\end{equation}

\begin{equation}
G=b\theta ^{\frac{3}{2}},
\end{equation}

\begin{equation}
\beta =\frac{c}{2}\theta ^{\frac{3}{2}}+c_{0}\theta ,
\end{equation}

\noindent where $a,b,c_{0}$ are constants.

\qquad Now it is convenient to put $\theta ^{\frac{1}{2}}=\rho .$ So, with
the help of (B.12), Eq. (B.6) can be written as

\begin{equation}
\nu \rho ^{2}f_{\rho \rho }+(\frac{1}{2}\rho ^{3}-c\rho ^{2}-2\nu \rho
)f_{\rho }+(c\rho +2\nu )f+a\rho ^{3}=0.
\end{equation}

The general solution of this equation is

\begin{equation}
f(\rho )=-2a\rho +\sqrt{\nu }\exp \frac{\rho (4c-\rho )}{4\nu }[c_{1}+c_{2}%
\func{erf}\sqrt{-\frac{(2c-\rho )^{2}}{4\nu }}].
\end{equation}

The form of $g(\rho )$ can be determined in the same way. Coming back to the
original variable $\theta $ and exploiting (B.1), (B.2), (B.11), the
solution (4.41a)--(4.41d) is readily found.

\bigskip \

\end{document}